\documentclass[12pt]{article}

\usepackage{comment}

\newcommand{\sect}{\sigma}
\newcommand{\dmsn}{n}

\usepackage[mathscr]{eucal}
\usepackage{amsmath,amsfonts,amssymb,amsthm}
\usepackage{times}
\usepackage{hyperref}

\numberwithin{equation}{section} 

\newtheorem{thm}{Theorem}[section]

 \newtheorem{prop}[thm]{Proposition}

\renewcommand{\tilde}{\widetilde}
\renewcommand{\hat}{\widehat}

\newcommand{\bref}[1]{\textbf{\ref{#1}}}

\newcommand{\gh}[1]{\mathrm{gh}(#1)}

\newcommand{\dx}{\mathrm{d}_X}

\newcommand{\derham}{\mathrm{d}}
\newcommand{\cM}{\mathcal{M}}

\renewcommand{\geq}{\,{\geqslant}\,}
\renewcommand{\leq}{\,{\leqslant}\,}

\newcommand{\binner}[2]{%
  {\langle}\kern-4.15pt{\langle}#1{,}\,#2{\rangle}\kern-4.15pt{\rangle}}

\newcommand{\ffrac}[2]{\raisebox{.5pt}%
  {\footnotesize$\displaystyle\frac{#1}{#2}$}\kern1pt}

\newcommand{\st}[2]{{\overset{#1}{#2}}}

\newcommand{\Liealg}{\mathfrak} 
\newcommand{\algg}{\Liealg{g}}

\newcommand{\fZ}{\mathbb{Z}}

 \def\cL{\mathcal{L}}
 \def\cN{\mathcal{N}}

\newcommand\blfootnote[1]{%
  \begingroup
  \renewcommand\thefootnote{}\footnote{#1}%
  \addtocounter{footnote}{-1}%
  \endgroup
}

\usepackage{ytableau,mathtools,leftindex}
\usepackage[numbers,sort&compress]{natbib}
 \usepackage[nottoc]{tocbibind}
\usepackage{authblk}
\makeatletter
\g@addto@macro\bfseries{\boldmath}
\makeatother

\usepackage{physics}

\title{Massive bigravity as a presymplectic BV-AKSZ sigma-model
}

\author[1,$\dagger$,$\ddagger$~~]{Maxim Grigoriev}

\author[2]{Vyacheslav~~Gritzaenko}

\affil[1]{\textsl{ Service de Physique de l'Univers, Champs et Gravitation, \protect\\ Universit\'e de Mons, 20 place du Parc, 7000 Mons, 
Belgium \vspace{5pt}}}
\affil[2]{\textsl{ Lebedev Physical Institute, \protect\\
  Leninsky ave. 53, 119991 Moscow, Russia \vspace{5pt}}}

\date{}

\begin{document}

\maketitle
\begin{abstract}
We propose a presymplectic BV-AKSZ sigma model encoding the ghost-free massive bigravity theory action as well as its Batalin-Vilkovisky extension in terms of the finite-dimensional graded geometry of the target space. A characteristic feature of the construction is that the target space is realised as a quasi-regular submanifold of a linear graded manifold which, in turn, is a direct product of two copies of the shifted Poincar\'e or (anti-)de Sitter Lie algebra. This graded manifold comes equipped with a natural presymplectcic structure and the compatible pre-$Q$ structure which is a sum of the Chevalley-Eilenberg differentials of each copy of the Lie algebra and the interaction term. The constraints determining the submanifold are the supergeometrical realisation of the known Deser-van Nieuwenhuizen condition and its descendant.
\end{abstract}
\vspace{2cm}

\noindent
\phantom{a}
\blfootnote{${}^{\dagger}$ Supported by the ULYSSE Incentive
Grant for Mobility in Scientific Research [MISU] F.6003.24, F.R.S.-FNRS, Belgium.}
\blfootnote{${}^{\ddagger}$ Also at Lebedev Physical Institute and Institute for Theoretical and Mathematical Physics, Lomonosov MSU.}
\tableofcontents

\section{Introduction}

Ghost-free massive bigravity~\cite{Hassan:2011zd,Paulos:2012xe} gives an example of the interacting theory of the massless graviton and its massive cousin. It was proposed as a natural generalisation of de Rham-Gabadadze-Tolley massive gravity (dRGT)~\cite{deRham:2010kj,deRham:2011rn}, to which it reduces upon setting one of the metrics to a fixed background value. This dRGT gravity theory gives the ghost-free interacting theory of massive graviton, resolving the earlier difficulties encountered in~\cite{vanDam:1970vg,Zakharov:1970cc,Boulware:1972yco}. 

Although the frame-like formulation~\cite{deRham:2014zqa} of the ghost-free massive bigravity (in what follows, this is just referred to as bigravity) is much less involved than the original metric-like one, the underlying geometrical structures still deserve thorough investigation. In particular, the origin of the algebraic constraints, including the Deser-van Nieuwenhuizen gauge condition~\cite{Deser:1974cy,Deffayet:2012zc,Ondo:2013wka} remains somewhat unclear. In this work we investigate the underlying geometrical structures from the point of view of differential-graded geometry and more specifically the Batalin-Vilkovisky (BV)~\cite{Batalin:1981jr,Batalin:1983wj} formalism.

The aforementioned BV formalism provides a general and very powerful framework to study interacting gauge theories, encoding the Lagrangian density and its gauge invariance inside the unique BV master action which satisfies the BV master equation. This gives rise to a number of natural homological complexes whose cohomology corresponds to crucial physical quantities such as observables, interactions, anomalies, global symmetries, etc., see e.g.~\cite{Barnich:2000zw} and references therein. Moreover, the BV approach is particularly useful for studying consistent interactions between gauge fields~\cite{Barnich:1993vg}.

Among possible BV formulations of local gauge theories, the minimal versions of so-called presymplectic BV-AKSZ sigma models are very concise and geometrical. In the case of topological theories this formulation reduces to the celebrated AKSZ sigma models~\cite{Alexandrov:1995kv} (see also~\cite{Cattaneo:1999fm,Grigoriev:1999qz,Batalin:2001fc,Cattaneo:2001ys,Roytenberg:2002nu,Barnich:2006hbb,Bonechi:2009kx,Barnich:2009jy,Bonavolonta:2013mza,Ikeda:2012pv} for further developments) whose target spaces are symplectic $Q$-manifolds. In the presymplectic BV-AKSZ approach~\cite{Alkalaev:2013hta,Grigoriev:2020xec,Grigoriev:2022zlq,Dneprov:2024cvt} (see also~\cite{Grigoriev:2016wmk,Dneprov:2022jyn}) all the data of a local gauge theory is encoded in the finite-dimensional geometry of the target space equipped with a possibly degenerate presymplectic structure and a compatible pre-$Q$ structure. Note that the target space is not necessarily finite-dimensional. For instance, it has to be infinite-dimensional in the presymplectic AKSZ formulation of interacting higher-spin theories~\cite{Sharapov:2021drr}, where the number of fields is infinite and interactions are higher-derivative.

In this work we identify a finite-dimensional graded geometry whose associated presymplectic BV-AKSZ formulation describes bigravity in three and four space-time dimensions. Although a presymplectic BV-AKSZ formulation of a given system can be constructed systematically starting from the jet bundle BV formulation and hence not usually of great interest, the specific version we propose here is not directly derivable from the usual formulation of bigravity, and moreover it possesses certain remarkable properties. In particular, it explicitly maintains the symmetry between the two sectors of bigravity. Namely, the underlying geometry is defined in terms of the ambient space which is a direct product of two copies of the shifted (anti-)de Sitter/Poincar\'e algebras and whose total BRST differential is deformed by the interaction term. In addition, the algebraic analogue of the Deser-van Nieuwenhuizen condition and its descendant is imposed, ensuring the right spectrum and gauge invariance of the theory. The associated presymplectic BV-AKSZ sigma model is then shown to describe correctly the BV formulation of bigravity.

The construction of this work gives a new insight into the geometrical structures underlying bigravity and, potentially, more general theories. The deformed and factorised structure of the graded manifold underlying this formulation suggests that other interesting theories could exhibit the same hidden structure. More concretely, this implies that new interacting models could be constructed immediately in the presymplectic BV-ASKZ form by deforming a direct product of the graded manifolds underlying two given gauge theories.
For instance, it may be possible to study massive deformations of supergravities in this way.

A characteristic feature of our construction is that the BV-AKSZ target space is effectively not a regular manifold but a singular surface in the linear purely-fermionic supermanifold. This is the price to pay for such a concise and symmetric formulation. The usual difficulties that arise with singular surfaces are avoided here since the target space is actually quasi-regular in the sense that its prolongation to the manifold of component superfields turns out to be regular. The same applies to the induced presymplectic structure that determines the conventional BV field-antifield space of the system. Note that this is not the first example of a singular target space in the AKSZ context: the recently proposed presymplectic BV-AKSZ formulation of the Plebanski gravity theory also employs a quasi-regular target space~\cite{Dneprov:2024cvt}. 

The paper is organised as follows: Section~\bref{sec:prelim} contains the necessary background material on massive bigravity as well as AKSZ sigma models and their presymplectic generalisations. The main Section is~\bref{sec:presymp-bigrav} where we introduce the presymplectic BV-AKSZ target space of bigravity in four dimensions, and we prove that the constraints determining the surface are quasi-regular. The analogous construction in three dimensions is also presented there. Some technical details are relegated to the Appendix.

\section{Preliminaries}
\label{sec:prelim}

\subsection{Bigravity in the frame-like approach}
\label{sec:frame-like}
The metric-like formulation of the ghost-free massive bigravity (bigravity, in what follows) is based on the following action~\cite{deRham:2014zqa}:
\begin{equation}
S[g,f]=\int d^4x(\kappa_1\sqrt{g}R_{\mu\nu}(g)g^{\mu\nu}+\kappa_2\sqrt{f}R_{\mu\nu}(f)f^{\mu\nu}+m^2\sqrt{g}U(\gamma))\,,
\end{equation}
where $g_{\mu\nu}$ and $f_{\mu\nu}$ are metric tensors, and symmetrical matrix $\hat{\gamma}$ is determined by $\gamma_\mu^\nu=(\sqrt{\hat{g}^{-1}\hat{f}})_\mu^\nu$, where $\hat{g}^{-1}$ and $\hat{f}$ 
are matrices with components $g^{\mu\nu}$ and $f_{\mu\nu}$, and $U(\gamma)=\beta_0+\beta_1Tr(\hat{\gamma})+\frac{1}{2}\beta_2(Tr(\hat{\gamma})^2-Tr(\hat{\gamma}^2))+\frac{1}{6}\beta_3(Tr(\hat{\gamma})^3-3Tr(\hat{\gamma}^2)Tr(\hat{\gamma}))+\beta_4(Tr(\hat{\gamma})^4-6Tr(\hat{\gamma}^2)Tr(\hat{\gamma})^2+3Tr(\hat{\gamma}^2)^2+8Tr(\hat{\gamma})Tr(\hat{\gamma}^3)-6Tr(\hat{\gamma}^4))$ for some constants $\beta_0,\dots,\beta_4$.

The structure of the Lagrangian becomes more clear when the theory is recast in the frame-like formulation, where the fundamental fields are the vierbeins $e_\mu^a$ and $f_\mu^a$. The respective action has the form~\cite{deRham:2014zqa}
\begin{equation}
\label{GFMBframe}
S[e,f]=\int\epsilon_{abcd}(\kappa_1R^{ab}(e)e^ce^d+\kappa_2R^{ab}(f)f^cf^d+m^2A^{abcd})\,,
\end{equation}
where $f_{\mu\nu}=\eta_{ab}f_\mu^af_\nu^b$, $R^{ab}(e)=\derham \omega^{ab}(e) +\omega^a{}_k(e)\omega^{kb}(e)$, $R^{ab}(f)=\derham \Tilde{\omega}^{ab}(f)+\Tilde{\omega}^a{}_k(f)\Tilde{\omega}^{kb}(f)$ and the pair of Lorentz connections $\omega$ and $\tilde\omega$ are expressed in terms of the respective vierbeins via the standard torsion-free conditions. Here and in what follows the wedge product of differential forms is assumed. The potential is given explicitly by:
\begin{equation}
\label{Abigrav}
A^{abcd}=C_0e^a e^b e^c e^d +C_2f^a f^b e^c e^d + C_4 f^a f^b f^c f^d\,.
\end{equation}

Note that one can consider more general $A^{abcd}$ which contains additional terms proportional to $feee$ and $efff$.  Nevertheless, \eqref{Abigrav} allows "physical" solution $C_0=\frac{1}{4}$, $C_2=-\frac{1}{2}$ and $C_4=\frac{1}{4}$, where "physical" means that (i) Minkowski space is a solution of EoMs of \eqref{GFMBframe} and (ii) $m^2$ is a square of mass for linearised theory. 
The above frame-like formulation is equivalent to the metric-like one provided the algebraic condition
$f_\mu^ae_{\nu a}=f_\nu^ae_{\mu a}$, i.e. $e^a f_a=0 $, is imposed on the vierbeins. This condition is known as Deser-Van Nieuwenhuizen gauge, see e.g.~\cite{Deffayet:2012zc,Ondo:2013wka}.  The first and the second terms of \eqref{GFMBframe} are invariant under diffeomorphisms with parameters $\varepsilon^\mu$ and $\Tilde{\varepsilon}^\mu$ respectively and local Lorentz transformations with parameters $\varepsilon^{ab}$ and $\Tilde{\varepsilon}^{ab}$ respectively. Nevertheless, the presence of the potential \eqref{Abigrav} decreases the symmetry of the action. 

The genuine first-order formulation of bigravity
is obtained by introducing auxiliary fields which in this case are coefficients of Lorentz connections in both sectors. The respective action reads as
\begin{equation}
\label{GFMBframe1o}
S[e,\omega, f,\Tilde{\omega}]=\int\epsilon_{abcd}(\kappa_1R^{ab}(\omega)e^ce^d+\kappa_2R^{ab}(\Tilde{\omega})f^cf^d+m^2A^{abcd})\,,
\end{equation}
and clearly reproduces \eqref{GFMBframe} upon the elimination of $\omega,\Tilde{\omega}$.

Let us briefly discuss the gauge invariance of the action~\eqref{GFMBframe1o}. The gauge transformations involving diffeomorphisms and local Lorentz transformations read as:
\begin{equation}
\label{GFMBgauge}
\begin{gathered}
\delta e^a = \derham(e_\mu^a\varepsilon^\mu)+\omega^{a}{}_ke^k_\mu\varepsilon^\mu+i_\varepsilon(\derham e^a+\omega^a_ke^k)-\varepsilon^{a}{}_be^b \\
\delta \omega^{ab} = \derham\varepsilon^{ab}+\omega^{a}{}_c\varepsilon^{cb}+\omega^{b}{}_c\varepsilon^{ac} \\
\delta f^a = \derham(f_\mu^a\varepsilon^\mu)+\Tilde{\omega}^{a}{}_kf^k_\mu\varepsilon^\mu+i_\varepsilon(\derham f^a+\Tilde{\omega}^a_ke^k)-\varepsilon^{a}{}_bf^b \\
\delta \Tilde{\omega}^{ab} = \derham\varepsilon^{ab}+\Tilde{\omega}^{a}{}_c\varepsilon^{cb}+\Tilde{\omega}^{b}{}_c\varepsilon^{ac}\,,
\end{gathered}
\end{equation}
where $\varepsilon^a=e^a_\mu\varepsilon^\mu$ and $\varepsilon^{ab}$ parameterize diffeomorphisms and local Lorentz transformations. On top of this, the theory has some additional algebraic gauge invariance which can be understood as a remnant of the off-diagonal local Lorentz transformations that survives the introduction of the interaction between the sectors. This is taken care of by an additional algebraic condition, namely, the Deser-Van Nieuwenhuizen condition $e^a f_a=0$, or in the component form: $f_{[\mu}^ae_{\nu] a}=0$.

Because $\omega$ and $\tilde\omega$ are auxiliary fields whose Euler-Lagrange equations express them in terms of the vierbeins, it is natural to immediately impose the extra condition constraining $\omega-\tilde\omega$. Namely,
\begin{equation}
(\omega^{ab}-\Tilde{\omega}^{ab}) e_a  f_b=0
\end{equation}
This is a consequence of $e_a f^a=0$ and the zero-torsion conditions $de^a+\omega^a{}_b e^b=0$ and $df^a+\tilde\omega^a{}_b f^b=0$ which are Euler-Lagrange equations for $\omega$ and $\tilde\omega$ respectively.

\subsection{AKSZ construction}

Let us briefly recall the AKSZ construction. The underlying geometrical object is a pair of $Q$-manifolds: target space $(\cM,Q)$ and source-space $(\mathcal{X},\delta_{\mathcal{X}})$. Recall that $Q$-manifold is a $\fZ$-graded supermanifold equipped with homological vector filed $Q$, i.e. a nilpotent vector field of degree $1$, $\gh{Q}=1$. The standard example of a $Q$-manifold is $T[1]X$, where $X$ is a real manifold. Its algebra of functions is the exterior algebra on $X$ and the $Q$-structure corresponds to the de Rham differential. We restrict ourselves to the case where $\mathcal{X}=T[1]X$ for a real space-time manifold of dimension $n$ and its $Q$-structire is the de Rham differential $\mathrm{d}_X$. We also assume that there are no physical fermions and hence the Grassmann parity $|\cdot|$ is  induced by $\fZ$-degree, i.e $|\cdot|=\gh\cdot\,\text{mod}\,2$.

Target space $(\cM,Q)$ is also equipped with the $Q$-invariant symplectic structure $\Omega$ of degree $n-1=\dim{X}-1$. It follows, there exists a Hamiltonian determined by:
\begin{equation}
    i_Q\Omega+d\cL=0\,.
\end{equation}
We also pick a symplectic potential $\chi$ such that $\Omega=d\chi$. Note that $\chi$ exists globally for $n>1$. This data is enough to define a gauge field theory whose action is defined as
\begin{equation}
\label{AKSZ-action}
    S[{\sect}]=\int_{T[1]X} {\sect}^*(\chi)(\derham_X)+{\sect}^*(\cL)\,,
\end{equation}
where $\sigma$ is a map $\sigma:T[1]X \to \cM$ (field configuration)
and $\sigma^*$ its associated pullback map which sends differential forms on $M$ to forms on $T[1]X$. The equations of motion determined by the action~\eqref{AKSZ-action}  read as
$\dx \circ \sigma^*=\sigma^* \circ Q$, i.e. imply that $\sigma$ is a $Q$-map. Moreover, this geometrical data also define the gauge symmetries of the above action.

It turns out that not only gauge transformations but also the complete  BV formulation of the system can be defined in terms of $(\cM,Q)$ and $(T[1]X,\dx)$.
More precisely, the fields and antifields arise as component superfields parameterizing the supermap $\hat\sigma$:
\begin{equation}
    \hat{\sect}^*(\Psi^A)=\st{0}{\Psi}{}^A+\st{1}{\Psi}{}^A{}_\mu{}\theta^\mu+\frac{1}{2}\st{2}{\Psi}{}^A{}_{\mu\nu}\theta^\mu\theta^\nu+\ldots\,.
\end{equation}
This field-antifield space is equipped with the symplectic structure which can be defined by giving its value on a pair of tangent vectors at a given point of the space of supermaps. More precisely, the point is preciesley the supermap $\hat\sigma$ while a tangent vector at $\hat\sigma$ is an infinitesimal variation $\delta\hat\sigma$. For  tangent vectors $\delta_1\hat\sigma$ and $\delta_2\hat\sigma$ at $\hat\sigma$ one sets
\begin{equation}
\bar\Omega_{\hat\sigma}(\delta_1\hat\sigma^*,\delta_2\hat\sigma^*)=
\int d^nx d^n\theta \Omega(x,\theta,\delta_1\hat\sigma^*(x,\theta),\delta_2\hat\sigma^*(x,\theta))\,,
\end{equation}
where $x^\mu,\theta^\mu$ are adapted coordinates on $T[1]X$. Finally, the BV master action extending \eqref{AKSZ-action} is given by:
\begin{equation}
\label{BV-AKSZ-action}
    S_{BV}[\hat{\sect}^*]=\int_{T[1]X} \hat{\sect}^*(\chi)(\derham_X)+\hat{\sect}^*(\cL)\,,
\end{equation}
It satisfies the classical BV master equation w.r.t. the above symplectic structure.

Despite being very elegant and concise the AKSZ construction is limited to topological systems unless $\cM$ is allowed to be infinite-dimensional or the source space is more general than $T[1](\text{spacetime})$. 

\subsection{Presymplectic BV-AKSZ and the space of superfields}

An interesting generalization of the AKSZ construction proposed in~\cite{Grigoriev:2022zlq}
(see also~\cite{Alkalaev:2013hta,Grigoriev:2016wmk,Grigoriev:2020xec,Dneprov:2022jyn} for particular cases and \cite{Dneprov:2024cvt} for the up-to-date exposition) is based on replacing the symplectic structure of the AKSZ construction with generically non-regular presymplectic structure and replacing the nilpotency condition on $Q$ by a weaker ``presymplectic'' master equation
\begin{equation}
\label{pmaster}
i_Qi_Q \Omega=0\,,
\end{equation}
which ensures that $Q$ is nilpotent on the symplectic quotient of $\cM$. Note that one still requires $Q$ to preserve the presymplectic structure,  $L_Q\Omega=0$, so that there exists $\cL$ such that $i_Q\Omega+d\cL=0$. Finally, the presymplectic structure on $\cM$ is required to be quasi-regular in the sense that its prolongation to the space of superfields is regular.

To make the regularity condition precise and to
give the presymplectic BV-AKSZ system unambiguous physical interpretation it is convenient to introduce the graded manifold $\bar \cM$ of AKSZ superfields. Namely, let us restrict to a local (in spacetime) analysis and pick a generic point $x \in X$. We then take $\bar \cM= \mathrm{SMaps}(T_x[1] X,\cM)$, i.e. a space of supermaps from $T_x[1]X$ to $\cM$, where $T_x[1]X=(T_xX)[1]$ is the tangent space at $x\in X$ with the degree shifted by $1$. Note that locally 
\begin{equation}
\mathrm{SMaps}(T[1]X,\cM)\cong \mathrm{SMaps}(X,\bar \cM)\,,
\end{equation}
see e.g.~\cite{Grigoriev:2020xec,Dneprov:2024cvt,Grigoriev:2024ncm} for more details. In other words the component superfields of the presymplectic AKSZ model are fields on $X$ taking values in $\bar\cM$. It is important to stress that we usually restricts ourselves to supermaps satisfying certain nondegeneracy condition.  In applications to gravity-like models we restrict to supermaps such that the frame fields are nondegenerate. 

Now the presymplectic structure $\Omega$ of degree $n-1$ determines a presymplectic structure on $\bar\cM$. In the coordinate terms it can be written as:
\begin{equation}
\Bar{\Omega}=\int d^\dmsn\theta\Omega_{AB}d\Psi^A(\theta)d\Psi^B(\theta)\,.
\end{equation}
The 2-form $\Omega$ is called quasi-regular if $\Bar{\Omega}$ is regular on $\bar\cM$ (recall that $\bar\cM$ is formed by supermaps satisfying nondegeneracy condition, otherwise nonregular presymplectic structure on $\cM$ determines a nonregular one 
 on the space of all supermaps). Given a regular $\bar\Omega$, one can define a symplectic quotient of $\bar\cM$, at least locally. Moreover, the restriction of the AKSZ-like  master action to component fields taking values in the symplectic quotient defines a BV system, giving the unambiguous physical interpretation to the presymplectic BV-AKSZ system, see \cite{Grigoriev:2022zlq,Grigoriev:2020xec,Dneprov:2024cvt} for details.
 
 The above version of the presymplectic BV-AKSZ formalism  is limited to diffeomorphism-invariant theories whose underlying bundles are globally trivial. A general version~\cite{Grigoriev:2022zlq,Dneprov:2024cvt} is formulated in terms of a fiber-bundle over $T[1]X$ where both $Q$ and $\Omega$, are defined on its total space $E$ replacing $\cM \times T[1]X$.


\section{Presymplectic BV-AKSZ formulation of bigravity}
\label{sec:presymp-bigrav}

In this section we construct the presymplectic BV-AKSZ formulation of bigravity. In contrast to the known examples of presympletic AKSZ systems, now the  target space is going to be a singular surface in the ambient linear supermanifold. However, this doesn't lead to problems because the prolongation of this surface to the space of supermaps turns out to be regular. We refer to such singular surfaces as to quasi-regular.

\subsection{Basic structures}

The ambient target space is the linear graded manifold $\cM$ with coordinates  $\xi^a$, $\rho^{ab}$, $\Tilde{\xi}^a$, $\Tilde{\rho}^{ab}$, all of them are of ghost degree 1 and $\rho^{ab}=-\rho^{ba}$ and $\Tilde{\rho}^{ab}=-\Tilde{\rho}^{ba}$. In more invariant terms, $\cM$ is a direct product of two copies of $\algg[1]$, where $\algg$ is a {Poincar\'e} or (anti) de Sitter ((A)dS) algebra, depending on the value of the parameters. Coordinates $\xi^a$ and $\Tilde{\xi}^a$ are associated to generators of translations (tranvections) and $\rho^{ab}$ and $\Tilde{\rho}^{ab}$ are associated to generators of Lorentz transformations. 
However,
the $Q$-structure on $\cM$ is not just a product $Q$-structure. Instead, it is the product $Q$-structure modified by an additional mixing term. 
In coordinates, $Q$ reads as follows:
\begin{equation}
\begin{gathered}
\label{Qbigrav}
Q\xi^a=\rho^a{}_k\xi^k \,,\\
Q\rho^{ab}=\rho^a{}_k\rho^{kb}+\frac{4C_0m^2}{\kappa_1}\xi^a\xi^b+\frac{2C_2m^2}{\kappa_1}\Tilde{\xi}^a\Tilde{\xi}^b \,,\\
Q\Tilde{\xi}^a=\Tilde{\rho}^a{}_k\Tilde{\xi}^k \,,\\
Q\Tilde{\rho}^{ab}=\Tilde{\rho}^a{}_k\Tilde{\rho}^{kb}+\frac{2C_2m^2}{\kappa_2}\xi^a\xi^b+\frac{4C_4m^2}{\kappa_2}\Tilde{\xi}^a\Tilde{\xi}^b\,.
\end{gathered}
\end{equation}
The $Q$-invariant presymplectic structure is just a sum of the respective presymplectic structures on the factors (strictly speaking, their pullbacks):
\begin{equation}
\label{presymp-total}
    \Omega=\epsilon_{abcd}(\kappa_1d\rho^{ab}d\xi^c\xi^d+\kappa_2d\Tilde{\rho}^{ab}d\Tilde{\xi}^c\Tilde{\xi}^d)=\Omega_{(e)}+\Omega_{(f)}\,.
\end{equation}

Note that for $C_2=0$ the above $Q$-manifold is a direct product of two copies of $\algg[1]$ seen as a presymplectic $Q$-manifolds with $Q$ structure being the Chevalley-Eilenberg differential of $\algg$ (seen as a vector field on $\algg[1]$) and the compatible presymplectic structure on $\algg[1]$ proposed in~\cite{Alkalaev:2013hta}.  The presymplectic AKSZ sigma-model with such target space $\algg[1]$ describes Einstein gravity~\cite{Alkalaev:2013hta,Grigoriev:2020xec} and hence the presymplectic AKSZ model with target-space $\cM$ define a direct product of two copies of Einstein gravity if $C_2=0$.

In the interesting case of $C_2\neq 0$, $Q$ is not nilpotent. Nevertheless, the presymplectic structure is $Q$-invariant so that the Hamiltonian 
satisfying $i_Q\Omega+d\cL=0$ exists and is given explicitly by:
\begin{multline}
    \cL= -\epsilon_{abcd}(\frac{1}{2}\kappa_1\rho^a{}_k\rho^{kb}\xi^c\xi^d+\frac{1}{2}\kappa_2\Tilde{\rho}^a{}_k\Tilde{\rho}^{kb}\Tilde{\xi}^c\Tilde{\xi}^d
    \\
+C_0 m^2\xi^a\xi^b\xi^c\xi^d+C_2m^2\Tilde{\xi}^a\Tilde{\xi}^b\xi^c\xi^d+C_4m^2\Tilde{\xi}^a\Tilde{\xi}^b\Tilde{\xi}^c\Tilde{\xi}^d)\,.
\end{multline}
Moreover, the presymplectic version $i_Qi_Q\Omega=0$ of the master-equation is satisfied provided we impose the following constraints:
\begin{equation}
\label{bgravconst}
   \xi^a\Tilde{\xi}_a=0
   \,, 
   \qquad 
     (\rho_{ab}-\Tilde{\rho}_{ab})\xi^a\Tilde{\xi}^b=0\,.
\end{equation}
This is shown in Appendix~\bref{app:master-eq}. 

The above conditions are $Q$-invariant in the sense that $Q$ preserves the ideal generated by the above constraints. Of course, these constraints are singular and the presymplectic master equation is to be understood algebraically as the condition that $i_Qi_Q\Omega$ belongs to the ideal generated by the left hand sides of~\eqref{bgravconst}. As we are going to see in the next section the prolongation of these constraints to the space of supermaps define a regular surface and hence a regular field theory. 

Finally, the AKSZ action~\eqref{AKSZ-action} determined by the above $Q,\Omega$ is precisely the frame like action \eqref{GFMBframe1o}. To make sure that the above data determines a proper BV formulation of the bigravity there remains to  check that (i) the prolongation of the constraints \eqref{bgravconst} to the space of superfields determine a regular surface, (ii) the presymplectic structure induced on the surface is regular, (iii)
the resulting BV action is proper, i.e. all the gauge invariances are properly taken into account. 
The last requirement is easily elucidated from the structure of the terms linear in ghosts of the BV-AKSZ action while the first two are checked in the next Section.  

\subsection{Regularity}
In this section we demonstrate that the prolongation of the constraints~\eqref{bgravconst} to $\bar\cM$ determines a regular submanifold $\bar \cN\subset \bar \cM$ and the presymplectic structure induced on the submanifold is regular, provided the allowed configurations for the frame fields $e_\mu^a$ and $f_\mu^a$ are such that 
$t_a^b=(e^{-1})^\mu_a f^b_\mu$ is sufficiently close to $\delta^a_b$. 

First of all let us see what happens when we set to zero all the coordinates on $\bar\cM$ of nonvanishing degree, i.e. restrict to submanifold $\bar\cM_0\subset \bar\cM$. Recall that $\cM_0$ is the space of maps (in contrast to supermaps) from $T_x[1]X$ to $\cM$. Introducing coordinates on $\bar\cM_0$ according to 
\begin{equation}
\begin{gathered}
\sect^*(\xi^a)=e_\mu^a\theta^\mu \,,\qquad
\sect^*(\rho^{ab})=\omega_\mu^{ab}\theta^\mu
\,,\\
\sect^*(\Tilde{\xi}^a)=f_\mu^a\theta^\mu
\,,\qquad
\sect^*(\Tilde{\rho}^{ab})=\Tilde{\omega}^{ab}_\mu\theta^\mu    
\end{gathered}
\end{equation}
the first condition in  \eqref{bgravconst} become $e^a{}_{\mu}f_{a\nu}-e^a{}_{\nu}f_{a\mu }=0$, reproducing the Deser-Van Nieuwenhuizen condition. 
The second condition from~\eqref{bgravconst} becomes $(\Tilde{\omega}_{ab}-\omega_{ab}) e^a f^b=0$ which is the additional condition introduced in Section~\bref{sec:frame-like}, so that the field content coincides.

Now we turn to fields of not necessarily vanishing degree. The component fields parameterizing supermap $\hat{\sigma}:T[1]X\to \cM$ are introduced as follows:
\begin{equation}
\label{coord-def1}
    \begin{gathered}
\hat{\sigma}^*(\xi^a)=\xi^a+e_\mu^a\theta^\mu+\frac{1}{2!}\xi^a_{\mu\nu}\theta^\mu\theta^\nu+\frac{1}{3!}\xi^a_{\mu\nu\lambda}\theta^\mu\theta^\nu\theta^\lambda+\ldots\,,\\
\hat{\sigma}^*(\rho^{ab})=\rho^{ab}+\omega_\mu^{ab}\theta^\mu+\frac{1}{2!}\rho^{ab}_{\mu\nu}\theta^\mu\theta^\nu+\frac{1}{3!}\rho^{ab}_{\mu\nu\lambda}\theta^\mu\theta^\nu\theta^\lambda+\ldots \,,
    \end{gathered}
\end{equation}
and 
\begin{equation}
\label{coord-def2}
    \begin{gathered}
\hat{\sigma}^*(\Tilde{\xi}^a)=\tilde\xi^a+f_\mu^a\theta^\mu+\frac{1}{2!}\tilde\xi^a_{\mu\nu}\theta^\mu\theta^\nu+\frac{1}{3!}\tilde\xi^a_{\mu\nu\lambda}\theta^\mu\theta^\nu\theta^\lambda+\ldots \,,\\
\hat{\sigma}^*(\Tilde{\rho}^{ab})=\tilde\rho^{ab}+\tilde\omega_\mu^{ab}\theta^\mu+\frac{1}{2!}\tilde\rho^{ab}_{\mu\nu}\theta^\mu\theta^\nu+\frac{1}{3!}\tilde\rho^{ab}_{\mu\nu\lambda}\theta^\mu\theta^\nu\theta^\lambda+\ldots \,,
    \end{gathered}
\end{equation}
where we slightly abuse notations by using the same notations  $\xi^a,\rho^{ab},\Tilde{\xi}^a,\Tilde{\rho}^{ab}$ to denote the respective superfield components. From the geometrical point of view $\hat\sigma$ is a supemap in contrast to the map $\sigma$.

The prolongations of the constraints are simply their pullbacks by $\hat{\sigma}^*$. By analyzing the prolongations of the constraints \eqref{bgravconst} to $\bar\cM$ order by order in $\theta$ one finds that they can be solved with respect to some of the coordinates provided the supermaps satisfy the conditions discussed above. More precisely: 
\begin{prop}
\label{prop:constr-reg}
Let us restrict to configurations such that $e_\mu^a$ and $f_\mu^a$ are invertible and $t_a^b=(e^{-1})^\mu_a f^b_\mu$ is sufficiently close to $\delta^a_b$. Then the prolongation of the constraints~\eqref{bgravconst} determine a regular surface $\bar\cN \subset \bar\cM$ and hence the constraints~\eqref{bgravconst} are quasi-regular. 
\end{prop}
\noindent
Details of the proof are given in  Appendix~\bref{app:constr-reg}. In particular, it follows from the analysis that the only remaining independent degree $1$-coordinates on $\bar\cN$ are $\xi^a$ and $\rho^{ab}$. In other words the off-diagonal diffeomorphisms and local Lorentz transformations do not survive so that the gauge invariance is exhausted by the genuine diffeomorphisms and diagonal local Lorentz transformations. These are obvious symmetries of the action \eqref{GFMBframe1o}. Note, however, that the BV formulation of the system arising from the above presymplectic formulation  encodes these gauge transformation in terms of the nonstandard generating set of gauge generators.

Let us now turn to the presymplectic structure \eqref{presymp-total} on $\cM$. Because it originates from the symplectic structures on the $e$ and $f$ factors the presymplectic structure it determines on $\bar\cM$ is regular. Indeed, for each of the factor the respective statement was proved in~\cite{Grigoriev:2020xec}. Now we use the analogous idea to demonstrate that $\bar\Omega$ remains regular when restricted to $\bar\cN \subset \bar\cM$.  

Denoting by $\bar\Omega_{\bar\cN}$ the restriction let us consider $\bar\Omega_{\bar\cN}$ at generic point of the body $\bar\cN_0 \subset \bar\cN$, i.e. where all the nonvanishing degree coordinates are set to zero. By picking a suitable coordinate system one can bring $\bar\Omega_{\bar\cN}$ at this point to the canonical form, which does not depend on the point and hence rank is constant along $\bar\cN_0$.
Moreover, the rank can only increase off $\bar\cN_0$ because the coordinate patch of $\bar\cN$ is a formal neighbourhood of its body. The analogous argument applies to the kernel distribution of
$\bar\Omega$, restricted to $\bar\cN$. Its rank can't decrease off $\bar\cN_0$ either but it is by construction a kernel distribution of $\bar\Omega|_{\bar\cN}$.  This considerations are straightforward generalisations of those from~\cite{Grigoriev:2020xec}. In this way we have arrived at the following:
\begin{prop}
Presymplectic structure $\bar\Omega_{\bar\cN}$ is regular.
\end{prop}

The above statements ensure that the proposed presymplectic BV-AKSZ formulation indeed determines a local BV formulation of bigravity. Indeed, if the target space was a regular surface the statement proved in~\cite{Grigoriev:2022zlq,Dneprov:2024cvt} (or a minor generalization of the one from~\cite{Grigoriev:2020xec}) ensures that the standard BV formulation emerges provided the presymplectic structure induced on the space of superfields (supermaps $T_x[1]X\to \cM$) is regular. Replacing the space of superfields with its regular submanifolds does not affect the proof.

\subsection{dRGT gravity through fixing background}

Bigravity is a generalization of dRGT gravity theory in the sense that the former reduces to the  latter if one sets fields of one of the sectors, say $\xi,\rho$, to their background values. More precisely, let $\bar e_{\mu}^a,\bar \omega_{\mu}^{ab}$  be a coefficients of a torsion-free  Poincar\'e connection describing gravitational background. Setting $e=\bar e, \omega=\bar\omega$ one finds that that the first term in the action \eqref{GFMBframe1o} together with the first term in \eqref{Abigrav} are fixed functions so that they can be disregarded. The resulting action takes the form: 
\begin{equation}
\label{drgt-action}
S[f,\tilde\omega|\bar e,\bar\omega]=\int\epsilon_{abcd}(R^{ab}(\tilde\omega)f^cf^d+m^2A^{abcd})\,,
\end{equation}
where we set $\kappa_2=1$ and
\begin{equation}
    A^{abcd}=C_4 f^a f^b f^c f^d +C_2f^a f^b \bar e^c \bar e^d\,,
\end{equation}
giving the action of dRGT gravity.

Note that setting $e,\omega$ to their background values breaks the diffeomorphism invariance down to the symmetries of the background (e.g. global Poincar\'e symmetry if $\bar e,\bar \omega$ describe flat Minkowski space)  while the local Lorentz invariance can be completely fixed by passing to the metric-like description. In other words the resulting theory is not a nontrivial gauge theory so that the  BV extension of the action~\eqref{drgt-action} is trivial and we do not dwell into it.

\subsection{Bigravity in 3 dimensions}

Now we show that essentially the same construction works for bigravity in 3 dimension. In this case the well-known frame like action reads as~\cite{Bergshoeff:2013xma} (see also \cite{Banados:2013fda,deRham:2012kf}):
\begin{multline}
\label{action3d}
    S[e^a,\omega^a]=\int (\kappa_1R^a(\omega^a)e_a+\kappa_2R^a(\Tilde{\omega})f_a+\\
    \epsilon_{abc}(C_0e^ae^be^c+C_1f^ae^be^c+C_2f^af^be^c+C_3f^af^bf^c))\,
\end{multline}
where $R_a=\epsilon_{abc}R^{bc}$ and $\omega_a=\epsilon_{abc}\omega^{bc}$, $\Tilde{\omega}_a=\epsilon_{abc}\Tilde{\omega}^{bc}$.

Just like in 4 dimensions, as a graded manifold $\cM$ we take a direct product of two copies of $\algg[1]$
where now $\algg$ is the Poincar\'e or (anti) de Sitter algebra in 3 dimensions. The total symplectic structure is again a sum of the symplectic structures of the factors, i.e.
\begin{equation}
    \Omega=\epsilon_{abc}(\frac{1}{2}\kappa_1d\rho^{ab}d\xi^c+\frac{1}{2}\kappa_2\Tilde{\rho}^{ab}d\Tilde{\xi}^c)\,.
\end{equation}
So far the construction is identical to the case of 4d except that the symplectic structure is now a true nondegenerate symplectic structure. This, of course, can easily be traced back to the Chern-Simons type formulation of gravity which is topological in 3d. 

The total pre-$Q$ structure again has the form of a sum of the Chevalley-Eilenberg  differentials of the factors deformed my the interaction term:
\begin{equation}
\begin{gathered}
Q\xi^a=\rho^a{}_k\xi^k \,,\\
Q\rho^{ab}=\rho^a{}_k\rho^{kb}+\frac{3C_0m^2}{\kappa_1}\xi^a\xi^b+\frac{C_2m^2}{\kappa_1}\Tilde{\xi}^a\Tilde{\xi}^b+\frac{C_1m^2}{\kappa_1}(\xi^a\Tilde{\xi}^b+\Tilde{\xi}^a\xi^b) \,,\\
Q\Tilde{\xi}^a=\Tilde{\rho}^a{}_k\Tilde{\xi}^k \,,\\
Q\Tilde{\rho}^{ab}=\Tilde{\rho}^a{}_k\Tilde{\rho}^{kb}+\frac{C_1m^2}{\kappa_2}\xi^a\xi^b+\frac{3C_3m^2}{\kappa_2}\Tilde{\xi}^a\Tilde{\xi}^b+\frac{C_2m^2}{\kappa_2}(\xi^a\Tilde{\xi}^b+\Tilde{\xi}^a\xi^b)\,.
\end{gathered}
\end{equation}
Note that for $C_1=C_2=0$ the above data defines the usual AKSZ sigma model with the target space $\cM$, which describes a product of two copies of 3d gravity with cosmological constants proportional to $C_0$ and $C_3$.

However, for $C_1$ and  $C_2$ nonvanishing the axioms of (presymplectic) BV-AKSZ formulation are not satisfied. 
Nevertheless, constraints \eqref{bgravconst} are again compatible with the constraints and ensure that $Q$ and $\Omega$ are compatible and the presymplectic master equation $i_Qi_Q\Omega=0$ is satisfied modulo the terms proportional to the constraints. More precisely,
\begin{equation}
    i_Qi_Q\Omega=2m^2\epsilon_{abc}\Sigma^c{}_k(C_2\Tilde{\xi}^a\Tilde{\xi}^b\xi^k-C_1\xi^a\xi^b\Tilde{\xi}^k)\,, \qquad  \Sigma^c{}_k=\rho^c{}_k-\Tilde{\rho}^c{}_k
\end{equation}
can be rewritten as
\begin{multline}
m^2\epsilon_{abc}(C_2\Tilde{\xi}^a\Tilde{\xi}^b\epsilon^{lck}\Sigma_l\xi_k-C_1\xi^a\xi^b\epsilon^{lck}\Sigma_l\Tilde{\xi}_k)=\\
2(m^2C_2\Tilde{\xi}^a\Tilde{\xi}^b\Sigma_{[b}\xi_{a]}
+m^2C_1\xi^a\xi^b\Sigma_{[a}\Tilde{\xi}_{b]})\,,
 \end{multline}
where $\Sigma_l=\epsilon_{lmn}\Sigma^{mn}$. It is easy to see that this expression is proportional to $\xi^a\tilde\xi_a$ which, in turn, is a first constraint of~\eqref{bgravconst}. In fact a similar approach can be employed to give an alternative proof of the analogous statement in 4d.

Finally, the Hamiltonian determined by $i_Q\omega+d\cL=0$ reads explicitly as
\begin{multline}
    \cL=-\epsilon_{abc}(\frac{1}{2}\kappa_1\rho^a{}_k\rho^{kb}\xi^c+\frac{1}{2}\kappa_2\Tilde{\rho}^a{}_k\Tilde{\rho}^{kb}\Tilde{\xi}^c+
    \\
m^2(C_0\xi^a\xi^b\xi^c+C_1\Tilde{\xi}^a\xi^b\xi^c+C_2\Tilde{\xi}^a\Tilde{\xi}^b\xi^c+C_3\Tilde{\xi}^a\Tilde{\xi}^b\Tilde{\xi}^c))\,,
\end{multline}
giving a representation of the frame like action~\eqref{action3d} in the presymplectic BV-AKSZ form.

\section{Conclusions}

The intricacy of ghost-free massive bigravity is characteristic of massive fields and their interactions. These difficulties can be traced to the lower-degree differential consequences of the Euler-Lagrange equations of massive fields, which are crucial in maintaining the number of degrees of freedom, see e.g.~\cite{Metsaev:2005ar,Zinoviev:2006im,Metsaev:2012uy,Kaparulin:2012px,Buchbinder:2012iz,deRham:2014zqa,Mazuet:2017hey,Mazuet:2018ysa,Grigoriev:2021wgw,Ochirov:2022nqz,Boulanger:2023lgd,Delplanque:2024xst,Delplanque:2024enh,Hinterbichler:2012cn,DelMonte:2016czb,Lust:2021jps,Lust:2023sfk,Kozuszek:2024vyb}.

The formulation of bigravity proposed in this work shows that constructing interactions for massive fields can be approached from the graded geometry perspective by employing quasi-regular surfaces in suitable graded spaces. In contrast to the standard approach where the free system is being deformed, our procedure can be interpreted as the introduction of interactions between two non-linear systems. This is accompanied by additional constraints which ensure the consistency of the resulting system by restricting both the fields and the gauge parameters. More specifically, one imposes constraints in the target space so that their prolongations restrict the gauge fields and parameters (ghosts). Note that the constraints also restrict higher components of the superfields, giving the correct spectrum of BV antifields.

Potential further developments are related to studying new consistent interactions between massive fields and, more generally, developing a general framework to analyse interactions involving massive fields within the presymplectic BV-AKSZ approach.

\section*{Acknowledgments}
\label{sec:Aknowledgements}
We appreciate discussions with A.~Mamekin, D.~Rudinsky,  Th.~Popelensky and especially I.~Dneprov. MG is grateful to E.~Skvortsov for useful discussions. The work of VG was supported by Theoretical Physics and Mathematics Advancement Foundation BASIS.

\appendix

\section{Presymplectic master equation}
\label{app:master-eq}

\def\txi{\tilde\xi}
In terms of  $\Sigma^{ab}=\rho^{ab}-\tilde\rho^{ab}$ the right hand side of the presymplectic master-equation~\eqref{pmaster} is proportional to:
\begin{equation}
\label{rhs-pme}
\epsilon_{abcd}\Sigma^a{}_k\txi^k\xi^b\xi^c\txi^d\,.
\end{equation}
It is convenient to employ the formalism of 2-component spinors~\cite{Penrose:1985bww,huggett1994introduction}. In this description $\xi^{[c}\txi^{d]}$ entering the above expression can be written as the decomposition into the selfdual and the anti-selfdual components:
\begin{equation}
\xi^{[c}\txi^{d]}=\epsilon^{C^\prime D^\prime}\Psi^{CD}+\epsilon^{CD}\bar\Psi^{C^\prime D^\prime}=\epsilon^{C^\prime D^\prime}\xi^C{}_{K^\prime}\txi^{DK^\prime}+\epsilon^{CD}\xi_K{}^{C^\prime}\txi^{KD^\prime}\,,
\end{equation}
where the indexes are raised and lowered by the antisymmetric tensor $\epsilon_{AB}$, $\epsilon_{12}=1$, 
$\epsilon^{AB}\epsilon_{CB}=\delta_C^A$ as $V_A=V^B\epsilon_{BA}, V^B=\epsilon^{BC}V_C$ and the 
contraction of the repeated indexes is assumed. Note that $\Psi^{CD}$ and $\bar\Psi^{C^\prime D^\prime}$ defined above are symmetric, i.e. $\Psi^{CD}=\Psi^{DC}$ and $\bar\Psi^{C^\prime D^\prime}=\bar\Psi^{D^\prime C^\prime }$, thanks to $\xi^a\txi_a=0$.

In a similar way, $\Sigma^{[a}{}_k\txi^k\xi^{b]}$ entering~\eqref{rhs-pme} can be written as:
\begin{multline}
    \Sigma^{[a}{}_k\txi^k\xi^{b]}=\epsilon^{A^\prime B^\prime}\Phi^{AB}+\epsilon^{AB}\bar\Phi^{A^\prime B^\prime}=\epsilon^{A^\prime B^\prime}(\Sigma^A{}_K\txi^{KL^\prime}\xi^B{}_{L^\prime}+\bar\Sigma^{L^\prime}{}_{K^\prime}\txi^{AK^\prime}\xi^B{}_{L^\prime})\\
    +\epsilon^{AB}(\Sigma_{LK}\txi^{KA^\prime}\xi^{LB^\prime}+\bar\Sigma^{A^\prime}{}_{K^\prime}\txi^{LK^\prime}\xi_L{}^{B^\prime})\,,
\end{multline}
 where $\bar\Sigma^{A^\prime B^\prime}$ and $\Sigma^{AB}$ denote the selfdual and the anti-selfdual components of $\Sigma^{ab}$. Note that the expressions in the parenthesis are symmetric in $AB$ and $A^\prime B^\prime$ respectively thanks to $\Sigma_{ab}\xi^a\txi^b=0$. 

The contribution of the anti-selfdual componenets to~\eqref{rhs-pme} is then proportional to
\begin{multline}
\Phi^{AB}\Psi_{AB}=(\Sigma^A{}_K\txi^{KL^\prime}\xi^B{}_{L^\prime}+\bar\Sigma^{L^\prime}{}_{K^\prime}\txi^{AK^\prime}\xi^B{}_{L^\prime})\xi_{AM^\prime}\txi_B{}^{M^\prime}=\\
=\Sigma_{AK}\txi^{KL^\prime}\xi^B{}_{L^\prime}\xi^{AM^\prime}\txi_{BM^\prime}+\bar\Sigma^{L^\prime}{}_{K^\prime}\txi^{AK^\prime}\xi^B{}_{L^\prime}\xi_{AM^\prime}\txi_B{}^{M^\prime}\,.
\end{multline}
The first term can be rewritten as
\begin{multline}
\Sigma_{AK}\txi^{KL^\prime}\xi^B{}_{L^\prime}\xi^{AM^\prime}\txi_{BM^\prime}=\Sigma_{AK}\txi^{KL^\prime}\xi_{BL^\prime}\xi^{BM^\prime}\txi^A{}_{M^\prime}=\\
=\Sigma_{AK}(\txi^{KL^\prime}\xi^C{}_{L^\prime})(\txi^{A M^\prime}\xi^{B}{}_{M^\prime})\epsilon_{BC}\,.
\end{multline}
Because $\Sigma_{AK}$ is symmetric while $\epsilon_{BC}$ is antisymmetric the expression vanishes. The same happens in the second term.

The analysis of the contribution $\bar\Phi^{A^\prime B^\prime}\bar\Psi_{A^\prime B^\prime}$ of the selfdual sectors is completely analogous and shows that it also vanishes. This shows that the master equation holds provided the constraints~\eqref{bgravconst} are imposed. 

\section{Regularity of $\bar{\cN}$}
\label{app:constr-reg}
Now we show that the surface $\bar\cN\subset \bar\cM$ determined by the the prolongation of conditions~\eqref{bgravconst} is regular under the conditions of the proposition. We employ the coordinate system on $\bar\cM$ determined by \eqref{coord-def1} and \eqref{coord-def2}. As the equations are algebraic it is enough to show the regularity at any submanifold singled out by the condition that $e_\mu^a$ is set to a particular value. By a linear transformation of the coordinates one can assume that $e_\mu^a=\delta_\mu^a$, this corresponds to defining component fields using $\theta^a=e_\mu^a\theta^\mu$ so that component fields are introduced as: 
\begin{equation}
\label{coord-def}
    \begin{gathered}
\hat{\sigma}^*(\xi_a)=\xi_a+\theta_a+\frac{1}{2!}\xi_{bc|a}\theta^b\theta^c+\ldots
\,,
\\
\hat{\sigma}^*(\txi_a)=\txi_a+f_{b|a}\theta^b+\frac{1}{2!}\txi_{bc|a}\theta^b\theta^c+\ldots
\\
\hat{\sigma}^*(\rho_{ab})=\rho_{ab}+\omega_{c|ab}\theta^c
+\frac{1}{2!}\rho_{cd|ab}\theta^c\theta^d+\ldots
\,,
\\
\hat{\sigma}^*(\tilde\rho_{ab})=\tilde\rho_{ab}+\tilde\omega_{c|ab}\theta^c
+\frac{1}{2!}\rho_{cd|ab}\theta^c\theta^d+\ldots\,,
    \end{gathered}
\end{equation}
where for future convenience we lowered the indexes using the Minkowski metric.

We seek for a solution for dependent components of $\hat\sigma^*(\txi_a)$
and $\hat\sigma^*\, (\tilde\rho_{ab})$ in the form of expansion in $\xi^a$ variables (to avoid confusion we mean the expansion in the $\theta^a$-independent  component superfield of $\hat\sigma^*(\xi^a)$). Then at each order in $\xi^a$ the prolongation of conditions~\eqref{bgravconst} can be uniquely solved for the totally-antisymmetric components of $\tilde\xi_{b_1\ldots b_l|a}$ and $\tilde\rho{}_{b_1\ldots b_l|ab}$, $0\leq l \leq 4$. It is convenient to denote by $(m,k)$ the component of \eqref{bgravconst} of degree $m$ in $\xi^a$ and degree $k$ in $\theta^a$. 
The notations for the homogeneous in $\xi^a$ components of $\tilde\xi_{b_1\ldots b_l|a}$ and $\tilde\rho{}_{b_1\ldots b_l|ab}$
are introduced as follows:
\begin{equation}
    \tilde\xi_{b_1\ldots b_l|a}=\Sigma_{k=0}^4{}^k\tilde\xi_{b_1\ldots b_l|a}\,,\qquad
    \tilde\rho_{b_1\ldots b_l|ab}=\Sigma_{k=0}^4{}^k\tilde\rho_{b_1\ldots b_l|ab}\,,
\end{equation}
where ${}^k\tilde\xi_{b_1\ldots b_l|a}$ and ${}^k\tilde\rho_{b_1\ldots b_l|ab}$ denote the respective components of order $k$ in $\xi^a$. We also keep using separate notation $f_{a|b}$ for  $\xi_{a|b}$.

Let us start with $\hat\sigma^*(\xi^a\txi_a)=0$ which is the prolongation of the first condition in~\eqref{bgravconst}. Its $(\bullet,1)$ component read as
\begin{equation}
    \xi^af_{b|a}-\txi_b=0\,.
\end{equation}
This is further decomposed into $(l,1)$ components
\begin{equation}
        (0,1):\,\, {}^0\txi_a=0\,, \qquad
        (l,1):\,\, {}^{l-1}f_{b|a}\xi^a-{}^l\txi_b=0\,
\end{equation}
which imply: ${}^0\txi^a=0$, ${}^l\txi_a={}^{l-1}f_{a|b}\xi^b$, $1\leq l\leq 4$.

Analogously, $(\bullet,2)$-equations read as
\begin{equation}
    \begin{gathered}
        (0,2): \xi^a_{mn}\,({}^0\txi_a)+f_{[m|m]}=0\,,\\
        (l,2): \xi^a_{mn}\,({}^l\txi_a)+{}^lf_{[m|n]}+\xi^a\,({}^{l-1}\txi_{mn|a})=0\,
    \end{gathered}
\end{equation}
and give ${}^0f_{[m|n]}=0$, ${}^lf_{[m|n]}=-\xi^a_{mn}({}^l\txi_a)-\xi^a({}^{l-1}\txi_{mn|a}$).
The considerations of $(\bullet,3)$-equation are completely analogous and show that they can be solved for ${}^k\txi_{[mn|p]}$, $0\leq k\leq 4$. The same is true for $(\bullet,4)$-equations which are solved for ${}^k\txi_{[mnp|q]}$.

In this way we have solved all the component equations except for $(\bullet,0)$ ones. It turns out that these are satisfied identically thanks to the $(\bullet,l)$, $1\leq l \leq 4$ equations. Indeed, $(0,0)$ equation is satisfied trivially. Then, $(l,0)$ equation can be written as ${}^{l-1}\txi_a\xi^a=0$. $(1,0)$ equation is satisfied because of ${}^{0}\txi_a=0$. The $(2,0)$ equation, 
\begin{equation}
    {}^0f_{a|b}\xi^a\xi^b=0\,,
\end{equation}
is satisfied thanks to ${}^0f_{a|b}={}^0f_{b|a}$. Analysis of  $(3,0)$ and $(4,0)$ equations is analogous but more involved and shows that they are also satisfied identically.

Let us turn to the second equation. First of all, for any antisymmetric tensor $\Sigma^\prime_{a_1\ldots a_k}$ let us introduce a map $\Phi_\lambda$ by the rule:
\begin{equation}
\Phi_\lambda (\Sigma^\prime_{a_1\ldots a_k})=\frac{1}{k}(\Sigma^\prime_{ca_2\ldots a_k} \lambda^c_{a_1}+\Sigma^\prime_{a_1c\ldots a_k} \lambda^c_{a_2}+\ldots+\Sigma^\prime_{a_1\ldots a_{k-1}c} \lambda^c_{a_k})\,.
\end{equation}
This is a linear map of the space of antisymmetric tensors to itself.  For $\lambda^a_b=\delta^a_b$ it is the identity map and hence $\Phi_\lambda$ is invertible (the inverse map will be denoted as $\Phi^{-1}_\lambda$). It must remain invertible for $\lambda_a^b$ sufficiently close to $\delta_a^b$ because the rank of the matrix can't be decreased by a small deformation. The condition that $\lambda^a_b$ is close to $\delta_a^b$
is satisfied by ${}^0f{}_{a|b}$ because in the basis where $e^a_b=\delta^a_b$ this is precisely the assumption of the Proposition~\bref{prop:constr-reg}.

Just like in the analysis of the first condition of \eqref{bgravconst} we decompose the prolongation of the second one into homogeneous components on homogeneity in $\xi^a$ and $\theta^a$ and denote by $(m,k)$ the respective homogeneous components. Moreover, it is convenient to introduce new variable ${\Sigma}_{c_1\ldots c_k|ab}=\rho_{c_1\ldots c_k|ab}-\tilde\rho_{c_1\ldots c_k|ab}$. Note, that the decomposition of $\tilde\rho_{c_1\ldots c_k|ab}$ with respect to $\xi^a$ induces the decomposition for ${\Sigma}_{\ldots|ab}$ and we denote by ${}^k{\Sigma}_{c_1\ldots c_k|ab}$ the $k$-th degree component.

It is convenient to start the analysis with $(0,k)$ equations, $2\leq k\leq4$, which have the following structure:
\begin{equation}
    \Phi({}^0\Sigma_{[\ldots|ab]})+\ldots=0\,, \qquad \Phi \equiv \Phi_{({}^0f)}
\end{equation}
These give: 
\begin{equation}
    \begin{gathered}
        {}^0\Sigma_{ab}=0\,, \qquad
        {}^0\Sigma_{[m|ab]}=0\,, \\
        {}^0\Sigma_{[mn|ab]}=-\Phi^{-1}({}^0\Sigma_{[m|ra}({}^0\txi^a_{nl]}))-\Phi^{-1}({}^0\Sigma_{[m|ab}({}^0f^a_n)\xi^b_{lr]})\,,
    \end{gathered}
\end{equation}
where antisymmetrisation in the last line is applied to $mnlr$ indexes.
It follows $(0,k)$ equations, $2\leq k\leq4$, can be solved for the totally-antisymmetric components of ${}^0\Sigma_{c_1\ldots c_{k-2}|ab}=0$. 

Furthermore, $(l,k)$ equations with  $l \geq 1,k\geq 2$ have a similar structure:
\begin{equation}
    \Phi({}^l\Sigma_{[\ldots|ab]})+\ldots=0\,,
\end{equation}
and are solved with respect for ${}^l\Sigma_{[c_1\ldots c_{k-2}|ab]}$ respectively. In particular, here we present explicitly the expressions for the lowest degree components: 
\begin{equation}
\label{bgcadd2}
    \begin{gathered}
        {}^1\Sigma_{mn}=\frac{1}{2}{}^0\Sigma_{b|mn}\xi^b+\frac{1}{2}\Phi^{-1}({}^0\Sigma_{a|mn}({}^0f^a_b)\xi^b)\,,\\
        {}^1\Sigma_{[m|nl]}=\frac{1}{2}({}^0\Sigma_{[mn|l]a}\xi^a+\Phi^{-1}({}^0\Sigma_{[mn|l]a}({}^1\txi^a))-\Phi^{-1}({}^0\Sigma_{[m|ab}({}^0\txi^a_{nl]})\xi^b)\\
        -\Phi^{-1}({}^0\Sigma_{[m|ab}({}^1\txi^a)\xi^b_{nl]})-\Phi^{-1}({}^1\Sigma_{a[l}{}^0\txi^a_{mn]})+\Phi^{-1}({}^1\Sigma_{ab}({}^0f^a_{[m})\xi^b_{nl]}))\,.
    \end{gathered}
\end{equation}

Finally, we are left with $(l,0)$ and $(l,1)$ equations. They are again satisfied identically provided the $(l,k)$ equations with  $l \geq 1,k\geq 2$ are satisfied. Indeed, $(0,0)$ equation is trivially satisfied. Equation $(1,0)$ reads explicitly as ${}^0\Sigma_{ab}({}^0\txi^a)\xi^b=0$ and is also satisfied identically thanks to ${}^0\Sigma_{ab}=0={}^0\txi^a$. The $(2,0)$ equation,
\begin{equation}
    {}^1\Sigma_{ab}({}^0\txi^a\xi^b)+{}^0\Sigma_{ab}({}^1\txi^a)\xi^b=0\,,
\end{equation}
is satisfied by the same reason. The $(3,0)$ equation has only one term that is not proportional to ${}^0\Sigma_{ab}$ or ${}^0\txi^a$, which is ${}^1\Sigma_{ab}{}^1\txi^a\xi^b$. But substituting ${}^1\Sigma_{ab}$ from \eqref{bgcadd2} and ${}^1\txi^a$ we obtain $\Phi_{({}^0f)}({}^0\Sigma_{[a|bc]})\xi^a\xi^b\xi^c$, which vanishes because ${}^0\Sigma_{[a|bc]}=0$. The analysis of the remaining component equations is analogous but more involved. Again, these equations  are also satisfied identically provided  the earlier considered component equations are satisfied. These complete the proof that $\bar\cN$ is regular.

\setlength{\itemsep}{0em}
\small
\providecommand{\href}[2]{#2}\begingroup\raggedright\endgroup

\end{document}